\documentclass[letterpaper,twocolumn,superscriptaddress,floatfix]{revtex4-2}

\usepackage{inputenc}
\usepackage{bm}
\usepackage{multirow,amssymb,amsbsy,amsmath}
\usepackage{graphicx}
\usepackage{float}
\usepackage{verbatim}
\makeatletter
\usepackage{pifont}
\usepackage{siunitx}
\usepackage{upgreek}
\usepackage{color}
\usepackage{indentfirst}
\usepackage{soul}
\usepackage{braket}
\usepackage[colorlinks=true,linkcolor=blue,citecolor=blue,urlcolor=blue]{hyperref}

\newcommand{\Rmnum}[1]{\expandafter\@slowromancap\romannumeral #1@}
\usepackage{etoolbox}
\usepackage{ragged2e}
\makeatletter
\apptocmd{\@caption}{\justifying}{}{}
\makeatother

\makeatletter
\renewcommand{\section}{\@startsection{section}{1}{0mm}
	{-\baselineskip}{0.5\baselineskip}{\bf\leftline}}
\makeatother

\begin{document}
\title{Room-Temperature Storage of Entanglement in a Silicon Carbide Quantum Node}

\author{Shuo Ren}
\thanks{These authors contributed equally to this work.}
\affiliation{Laboratory of Quantum Information, University of Science and Technology of China, Hefei, Anhui 230026, China}
\affiliation{Anhui Province Key Laboratory of Quantum Network, University of Science and Technology of China, Hefei, Anhui 230026, China}
\affiliation{CAS Center For Excellence in Quantum Information and Quantum Physics, University of Science and Technology of China, Hefei, Anhui 230026, China}
\affiliation{Hefei National Laboratory, University of Science and Technology of China, Hefei, Anhui 230088, China}

\author{Rui-Jian Liang}
\thanks{These authors contributed equally to this work.}
\affiliation{Laboratory of Quantum Information, University of Science and Technology of China, Hefei, Anhui 230026, China}
\affiliation{Anhui Province Key Laboratory of Quantum Network, University of Science and Technology of China, Hefei, Anhui 230026, China}
\affiliation{CAS Center For Excellence in Quantum Information and Quantum Physics, University of Science and Technology of China, Hefei, Anhui 230026, China}
\affiliation{Hefei National Laboratory, University of Science and Technology of China, Hefei, Anhui 230088, China}

\author{Qi-Cheng Hu}
\affiliation{Laboratory of Quantum Information, University of Science and Technology of China, Hefei, Anhui 230026, China}
\affiliation{Anhui Province Key Laboratory of Quantum Network, University of Science and Technology of China, Hefei, Anhui 230026, China}
\affiliation{CAS Center For Excellence in Quantum Information and Quantum Physics, University of Science and Technology of China, Hefei, Anhui 230026, China}
\affiliation{Hefei National Laboratory, University of Science and Technology of China, Hefei, Anhui 230088, China}

\author{Zhen-Xuan He}
\affiliation{Laboratory of Quantum Information, University of Science and Technology of China, Hefei, Anhui 230026, China}
\affiliation{Anhui Province Key Laboratory of Quantum Network, University of Science and Technology of China, Hefei, Anhui 230026, China}
\affiliation{CAS Center For Excellence in Quantum Information and Quantum Physics, University of Science and Technology of China, Hefei, Anhui 230026, China}
\affiliation{Hefei National Laboratory, University of Science and Technology of China, Hefei, Anhui 230088, China}
   
\author{Ji-Yang Zhou}
\affiliation{Laboratory of Quantum Information, University of Science and Technology of China, Hefei, Anhui 230026, China}
\affiliation{Anhui Province Key Laboratory of Quantum Network, University of Science and Technology of China, Hefei, Anhui 230026, China}
\affiliation{CAS Center For Excellence in Quantum Information and Quantum Physics, University of Science and Technology of China, Hefei, Anhui 230026, China}
	
\author{Wu-Xi Lin}
\affiliation{Laboratory of Quantum Information, University of Science and Technology of China, Hefei, Anhui 230026, China}
\affiliation{Anhui Province Key Laboratory of Quantum Network, University of Science and Technology of China, Hefei, Anhui 230026, China}
\affiliation{CAS Center For Excellence in Quantum Information and Quantum Physics, University of Science and Technology of China, Hefei, Anhui 230026, China}
\affiliation{Hefei National Laboratory, University of Science and Technology of China, Hefei, Anhui 230088, China}	
   
\author{Zhi-He Hao}
\affiliation{Laboratory of Quantum Information, University of Science and Technology of China, Hefei, Anhui 230026, China}
\affiliation{Anhui Province Key Laboratory of Quantum Network, University of Science and Technology of China, Hefei, Anhui 230026, China}
\affiliation{CAS Center For Excellence in Quantum Information and Quantum Physics,
University of Science and Technology of China, Hefei, Anhui 230026, China}

\author{Tao Tu}
\affiliation{Laboratory of Quantum Information, University of Science and Technology of China, Hefei, Anhui 230026, China}
\affiliation{Anhui Province Key Laboratory of Quantum Network, University of Science and Technology of China, Hefei, Anhui 230026, China}
\affiliation{CAS Center For Excellence in Quantum Information and Quantum Physics, University of Science and Technology of China, Hefei, Anhui 230026, China}
\affiliation{Hefei National Laboratory, University of Science and Technology of China, Hefei, Anhui 230088, China}

\author{Jin-Shi Xu}
\altaffiliation{Email: jsxu@ustc.edu.cn}
\affiliation{Laboratory of Quantum Information, University of Science and Technology of China, Hefei, Anhui 230026, China}
\affiliation{Anhui Province Key Laboratory of Quantum Network, University of Science and Technology of China, Hefei, Anhui 230026, China}
\affiliation{CAS Center For Excellence in Quantum Information and Quantum Physics, University of Science and Technology of China, Hefei, Anhui 230026, China}
\affiliation{Hefei National Laboratory, University of Science and Technology of China, Hefei, Anhui 230088, China}

\author{Chuan-Feng Li}
\altaffiliation{Email: cfli@ustc.edu.cn}
\affiliation{Laboratory of Quantum Information, University of Science and Technology of China, Hefei, Anhui 230026, China}
\affiliation{Anhui Province Key Laboratory of Quantum Network, University of Science and Technology of China, Hefei, Anhui 230026, China}
\affiliation{CAS Center For Excellence in Quantum Information and Quantum Physics, University of Science and Technology of China, Hefei, Anhui 230026, China}
\affiliation{Hefei National Laboratory, University of Science and Technology of China, Hefei, Anhui 230088, China}
 
\author{Guang-Can Guo}
\affiliation{Laboratory of Quantum Information, University of Science and Technology of China, Hefei, Anhui 230026, China}
\affiliation{Anhui Province Key Laboratory of Quantum Network, University of Science and Technology of China, Hefei, Anhui 230026, China}
\affiliation{CAS Center For Excellence in Quantum Information and Quantum Physics, University of Science and Technology of China, Hefei, Anhui 230026, China}
\affiliation{Hefei National Laboratory, University of Science and Technology of China, Hefei, Anhui 230088, China}

\begin{abstract}
Robust entanglement at room temperature is a central challenge for solid-state quantum information processing and quantum-enhanced sensing. Here we demonstrate room-temperature storage of entanglement in a silicon carbide (SiC) quantum node by coherently transferring an electron--nuclear entangled state onto long-lived nuclear-spin memory qubits. Using a shallow single PL6 color center in 4H--SiC, we realize a fully addressable three-qubit register composed of one electron-spin processor and two strongly coupled $^{29}$Si nuclear-spin memory qubits. This platform enables the deterministic generation of high-fidelity entangled states, including a nuclear-spin Bell state with a fidelity of $94 \pm 2~\%$ and a three-qubit GHZ-type state with a fidelity of $89 \pm 4~\%$. By implementing a SWAP-gate protocol in the strong hyperfine-coupling regime, the electron--nuclear entanglement is transferred to the nuclear-spin memory with a fidelity of $92.5 \pm 2.5~\%$, extending the entanglement lifetime by a factor of 240. We further confirm the generality of this approach in an additional heterogeneous $^{29}$Si--$^{13}$C nuclear-spin register and, through a statistical survey of 200 single PL6 centers, show that multi-nuclear-spin registers occur naturally with probabilities above 10\%. These results position shallow SiC color centers as a powerful platform for entanglement-assisted quantum sensing and scalable room-temperature quantum technologies.
\end{abstract}

\date{\today}
\maketitle

Entanglement is a key resource for quantum-enhanced sensing and quantum information processing, enabling capabilities beyond classical limits \cite{Horodecki2009,Rovny2025,Zhou2025}. In solid-state platforms, however, entangled states are highly susceptible to environmental decoherence, which severely limits their operation under ambient conditions \cite{Schlosshauer2005,Chirolli2008}. This challenge becomes particularly pronounced for near-surface spin defects, which are of interest for a broad range of quantum technologies owing to their strong coupling to the surrounding environment and compatibility with nanoscale device architectures. While shallow defects offer enhanced accessibility and integration potential, they are also exposed to increased magnetic, electric, and surface-related noise, leading to reduced electronic-spin coherence times~\cite{Rosskopf2014,Myers2014,Kim2015,Li2025}. As a consequence, even when entanglement can be generated, its lifetime is often too short to be practically useful. Addressing this limitation requires strategies that protect fragile entangled states against decoherence without compromising the advantages intrinsic to shallow-defect platforms.

A promising approach is to exploit hybrid electron--nuclear spin systems, in which a fast and optically addressable electron spin is coherently coupled to nearby long-lived nuclear spins \cite{Dutt2007Science,Maurer2012,Bradley2019,Grimm2025,Dolde2013}. 
In such systems, nuclear spins can serve as robust quantum memories, allowing fragile quantum states—including entangled states—to be transferred away from decoherence-prone electronic degrees of freedom and preserved for extended durations. Demonstrating coherent entanglement transfer and storage in such hybrid registers is therefore a key step toward long-lived multi-qubit quantum control in solid-state systems.

Optically addressable color centers in silicon carbide (SiC) offer a particularly attractive platform for realizing such hybrid spin registers~\cite{Klimov2015,Bourassa2020,Lai2024,Hu2024,Hesselmeier2024,Ren2026}. In particular, shallow SiC defects can combine robust spin--optical properties with surface accessibility, which is essential for sensing-compatible quantum registers~\cite{Li2025}. Its wafer-scale growth, deterministic defect creation, and compatibility with established micro- and nanofabrication techniques enable scalable device integration \cite{Koehl2011,Christle2015,Widmann2015,Seo2016,Christle2017,Anderson2019,Nagy2019,Wang2020,Hu2026AdvMater,Crook2020,Babin2022,Son2022,Castelletto2022,Li2022,Zhou2023,He2024,Zhou2025APR}. These features make SiC well suited for near-surface quantum architectures.

Among the various SiC color centers, the PL6 center hosting an S=1 ground-state triplet in 4H--SiC stands out due to its robust room-temperature spin properties, favorable optical addressability, and strong hyperfine coupling to nearby nuclear spins \cite{Li2022}. These strong couplings enable coherent electron--nuclear spin interactions~\cite{Hu2024,Ren2026} and provide access to multi-qubit registers in the immediate vicinity of the surface. Nevertheless, the electronic-spin coherence of shallow PL6 centers remains limited, posing a bottleneck for preserving entanglement over extended timescales. Whether and how entanglement can be protected in such shallow, room-temperature solid-state environments therefore constitutes an important open question.

In this work, we demonstrate room-temperature protection of entanglement in silicon carbide by coherently transferring an electron--nuclear entangled state associated with a single near-surface PL6 center onto two strongly coupled $^{29}$Si nuclear spins. Using a fully addressable three-qubit register, we deterministically prepare electron--nuclear entangled states and implement a SWAP-gate-based protocol to map the entanglement onto the nuclear-spin subsystem. This conversion transforms a fragile electron--nuclear entangled state into a long-lived nuclear-spin entangled memory, extending the entanglement lifetime by up to a factor of 240 while maintaining high fidelity under ambient conditions. 

The generality and reproducibility of this protocol have been further validated by demonstrating entanglement storage in a distinct heteronuclear $^{29}$Si--$^{13}$C register and by performing a statistical survey of 200 single PL6 centers fabricated using two independent implantation methods. The survey shows that multi-nuclear-spin registers containing two or more strongly coupled nuclear spins occur naturally with probabilities above 10\%. These results further show that suitable multi-spin resources arise naturally in SiC without isotopic engineering or post-selection of exceptionally rare defects.

\begin{figure}[htbp]
\centering
\includegraphics[scale = 0.8]{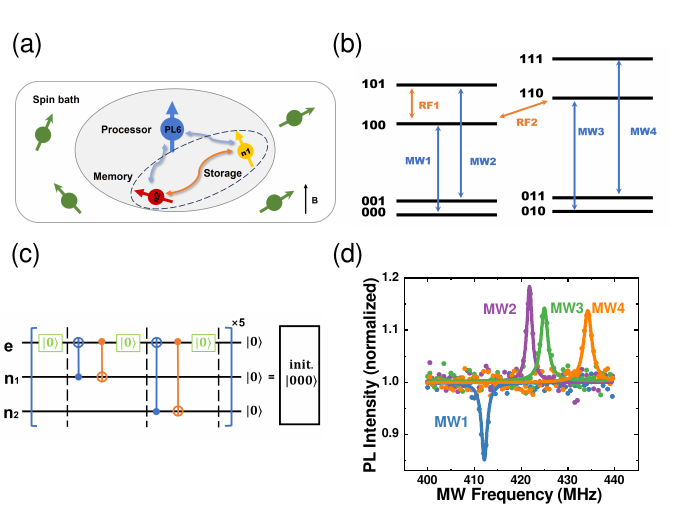}
  \caption{Energy levels of a PL6 center coupled to two $^{29}$Si nuclear spins in silicon carbide. (a) Schematic of a PL6 electron spin coupled to two proximal nuclear spins forming a three-qubit register, with the electron spin as a processor and the nuclear spins as memory qubits. (b) Simplified energy-level diagram showing the eight computational basis states $\ket{000}$–$\ket{111}$. Blue arrows indicate electron-spin transitions conserving nuclear spin, while orange arrows denote nuclear-spin–flip transitions associated with different hyperfine couplings. (c) Pulse sequence for initializing the three-qubit register by transferring electron-spin polarization to the nuclear spins. (d) Pulsed ODMR spectrum after initialization, confirming individual nuclear-spin preparation; solid curves are theoretical predictions for the MW1–MW4 transitions.
 }
\label{Figure 1}
\end{figure}

Our silicon carbide quantum node is based on a shallow single PL6 center in 4H–SiC, whose electron spin acts as a fast processing qubit and is strongly coupled to two nearby $^{29}$Si nuclear spins that serve as long-lived memory qubits (Fig.~\ref{Figure 1}(a)). This hybrid electron–nuclear architecture combines the rapid control and optical accessibility of an electron spin with the superior coherence properties of nuclear spins, forming a minimal yet powerful multi-qubit register capable of storing quantum correlations well beyond the electron-spin coherence time at room temperature.

In the experiment, shallow PL6 centers were created by N$_2^+$ ion implantation. According to Stopping and Range of Ions in Matter (SRIM) simulations, the implantation profile produces defects with an expected peak depth of approximately 15~nm below the surface. The fluorescence collection efficiency was further enhanced using a plasmon-assisted optical configuration~\cite{Zhou2023} (see SM Note 1~\cite{Supplemental} for details). The assignment of the investigated defects to PL6 centers is confirmed by low-temperature photoluminescence spectroscopy and zero-field ODMR measurements, which show the characteristic PL6 zero-phonon line near 1038~nm and a zero-field splitting around 1351~MHz~\cite{Li2022,Falk2013NatCommun}, respectively (see SM Note 2~\cite{Supplemental} for details). All spin-control and entanglement-storage measurements were performed at room temperature.

Under a static magnetic field of $B=330$~G aligned along the defect symmetry axis, the Zeeman interaction lifts the degeneracy of the $m_s=\pm1$ states. Throughout this work, we encode the electron-spin qubit in the ${m_s=0,-1}$ subspace, which enables robust optical initialization and efficient microwave manipulation. The host 4H–SiC crystal possesses a natural isotopic composition of $\sim1.1\%$ $^{13}$C and $\sim4.7\%$ $^{29}$Si nuclear spins ($I=1/2$). The two memory qubits are associated with nearby $^{29}$Si nuclear spins, which are coupled to the electron spin via anisotropic hyperfine interactions.

Together, the electron spin and two nuclear spins form an eight-level quantum register described in the computational basis $\ket{m_s, m_{I_1}, m_{I_2}}$, spanning states $\ket{000}$ through $\ket{111}$ (Fig.~\ref{Figure 1}(b)). The two nuclear spins occupy inequivalent lattice sites in 4H--SiC and are assigned to the $\mathrm{Si_{IIb}}$ ($n_{1}$) and $\mathrm{Si_{IIa}}$ ($n_{2}$) configurations based on their measured hyperfine splittings~\cite{Falk2015}. The corresponding longitudinal hyperfine coupling constants are $A_{zz}^{(n_1)}=9.6$~MHz and $A_{zz}^{(n_2)}=12.5$~MHz, respectively. The resulting distinct hyperfine coupling strengths produce spectrally resolvable transitions in ODMR, enabling selective control of each nuclear spin. In the present magnetic-field regime, direct nuclear--nuclear interactions are negligible, and the system dynamics are governed by the individual hyperfine couplings to the electron spin. This hierarchy of interaction scales allows the electron spin to act as a coherently addressable quantum bus, mediating conditional gate operations and coherent state transfer between the processor electron spin and long-lived nuclear-spin memory qubits. The coupled electron--nuclear register is well described by a standard hyperfine Hamiltonian, whose parameters are extracted from ODMR spectroscopy. Complete details of the Hamiltonian model, hyperfine parameters, and parameter determination are provided in SM Note 2~\cite{Supplemental}.

\begin{figure}[htbp]
\centering
\includegraphics[scale = 0.8]{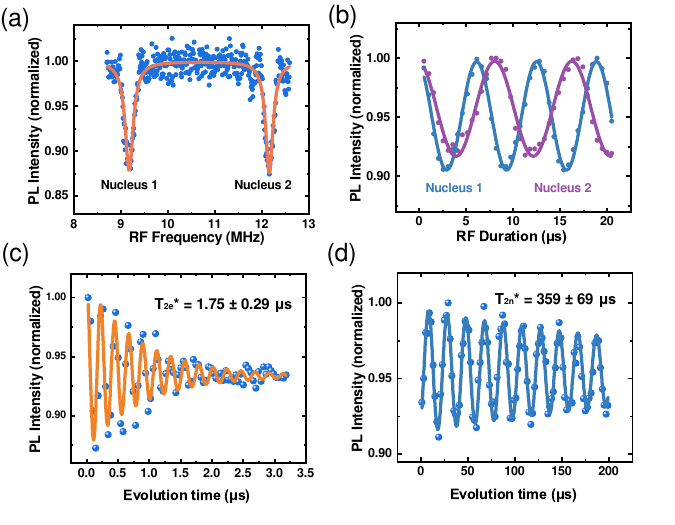}
  \caption{Coherent control of a three-qubit register composed of a PL6 electron spin and two $^{29}$Si nuclear spins. 
(a) Optically detected nuclear magnetic resonance spectrum showing two resolved resonances at 12.16 and 9.18 ~MHz, corresponding to nuclear-spin transitions of two inequivalent $^{29}$Si nuclei (labeled nucleus 1 and nucleus 2), each hyperfine-coupled to the electron spin of a single PL6 center. (b) Rabi oscillations of the two nuclear spins $n_1$ and $n_2$ at $B=330$ G with the electron spin prepared in $m_s=-1$. (c) Ramsey fringes of the electron spin. (d) Ramsey fringes of nuclear spin $n_2$; solid lines are fits to the data.
} 
\label{Figure 2}
\end{figure}

Reliable generation and storage of entanglement require deterministic initialization and coherent control of all qubits in the hybrid register. We implement a tailored sequence of microwave (MW) and radio-frequency (RF) pulses that repeatedly transfers polarization from the optically initialized electron spin to the nuclear-spin memory qubits (Fig.~\ref{Figure 1}(c)), thereby preparing the system in the $\ket{000}$ state with high fidelity at room temperature. 

The effectiveness of the initialization protocol is verified by pulsed ODMR measurements. After initialization, only the transitions corresponding to the prepared nuclear-spin configuration are observed, while all other transitions are strongly suppressed (Fig.~\ref{Figure 1}(d)). This selective spectral response provides direct evidence that both nuclear spins are polarized into well-defined states and that the full three-qubit register is initialized deterministically. Further details on the pulse sequences and analysis of the system polarization are provided in SM Note 3~\cite{Supplemental}.

Following initialization, we characterize the coherent control of the nuclear-spin qubits. The nuclear-spin transitions are addressed by applying resonant radio-frequency fields and monitored through the electron-spin-dependent optical signal, a protocol known as optically detected nuclear magnetic resonance (ODNMR)~\cite{Klimov2015}. ODNMR spectroscopy reveals two well-resolved resonances at frequencies of 9.18~MHz and 12.16~MHz (Fig.~\ref{Figure 2}(a)), corresponding to two inequivalent $\mathrm{Si_{IIb}}$ and $\mathrm{Si_{IIa}}$ nuclear spins, respectively, in good agreement with previous results~\cite{Falk2015}. The small difference between the ODNMR resonance frequencies and the hyperfine splittings extracted from ODMR arises from the finite nuclear Zeeman contribution at the applied magnetic field. Owing to their long coherence times, the ODNMR linewidths are much narrower than those observed in ODMR. Moreover, the spectral isolation of the resonances enables frequency-selective addressing of the individual nuclear spins $n_1$ and $n_2$ with negligible crosstalk. Using this register-specific addressability, we apply tailored radio-frequency pulse sequences to drive nuclear Rabi oscillations, as shown in Fig.~\ref{Figure 2}(b). The blue and purple fitted curves correspond to the coherent oscillations of the individual nuclear spins $n_1$ and $n_2$, respectively. More precisely, the observed oscillations arise from the transitions between $\ket{100} \leftrightarrow \ket{101}$ and $\ket{100} \leftrightarrow \ket{110}$, demonstrating selective control within the electron-spin-resolved nuclear subspaces. Rabi oscillations associated with other nuclear-spin transitions are presented in SM Note 3~\cite{Supplemental}.

To benchmark the memory advantage of the nuclear spins relative to the electron spin, we measure the coherence properties of the individual qubits using Ramsey interferometry. The electron spin exhibits an inhomogeneous dephasing time of $T_{2e}^* = 1.75 ~\pm ~0.29~\upmu\mathrm{s}$ (Fig.~\ref{Figure 2}(c)), consistent with room-temperature operation in a natural SiC crystal~\cite{Li2022,Zhou2023,Hu2024}. The Ramsey interferometry result for nuclear spin $n_2$ is presented here to characterize the nuclear-spin memory coherence, while the corresponding measurement for $n_1$ is provided in SM Note 3~\cite{Supplemental}. The nuclear spin exhibits a much longer coherence time of $T_{2n}^* = 359 ~\pm ~69~\upmu\mathrm{s}$ (Fig.~\ref{Figure 2}(d)), exceeding the electron-spin coherence time by more than two orders of magnitude. 

\begin{figure}[htbp]
\centering
\includegraphics[scale = 0.8]{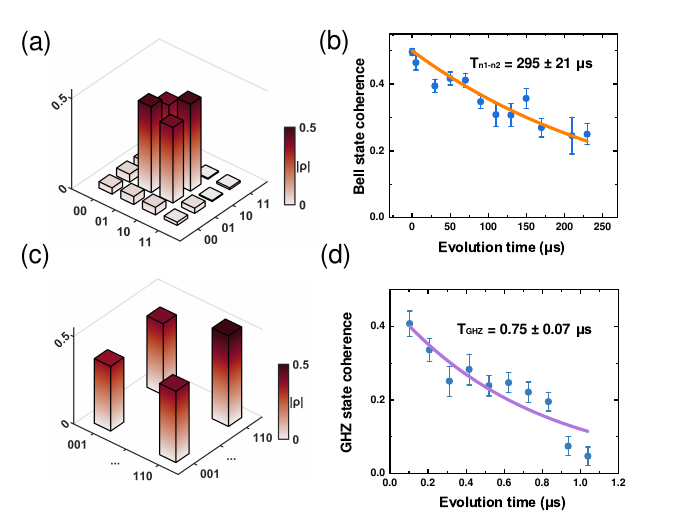}
\caption{Entanglement of two nuclear spins and a three-qubit GHZ-type state. 
(a) Reconstructed density matrix of the nuclear-spin Bell state obtained from quantum state tomography. 
(b) Measured coherence decay of the nuclear-spin Bell state. 
(c) Reconstructed dominant density-matrix elements of the three-qubit GHZ-type state. 
(d) Measured coherence decay of the GHZ-type state. Error bars denote statistical uncertainties.
}
\label{Figure 3}
\end{figure} 

With full initialization and coherent control of the three-qubit register established, we proceed to generate entangled states involving the nuclear spins and the complete electron--nuclear system. Starting from the initialized state $\ket{000}$, entanglement is generated using sequences of MW and RF pulses that exploit the strong hyperfine coupling between the electron spin and the nuclear spins. The pulse sequence implements a controlled-phase gate on the electron spin which, combined with appropriately timed $\pi/2$ rotations on the nuclear spins, effectively realizes a nuclear-spin controlled-NOT (CNOT) operation~\cite{Waldherr2014Nature}. Using this approach, we first generate a maximally entangled Bell state of the two nuclear spins,
\[
\frac{1}{\sqrt{2}}\left(\ket{101} + \ket{110}\right),
\]
in which the electron spin remains disentangled from the nuclear-spin subsystem. The detailed derivation of the quantum state is provided in SM Note 3~\cite{Supplemental}. Quantum state tomography is performed to reconstruct the density matrix of the nuclear-spin pair, yielding a state fidelity of $94 \pm 2~\%$ with respect to the ideal Bell state (Fig.~\ref{Figure 3}(a)). The reconstructed density matrix exhibits strong off-diagonal coherence and negligible population outside the target subspace, demonstrating high-quality preparation of nuclear-spin entanglement at room temperature. Details of the quantum state tomography procedure, including the applied pulse sequences and the analysis of fidelity errors, are provided in SM Note 7~\cite{Supplemental}.

We further extend this protocol to generate a three-qubit GHZ-type entangled state of the form
\[
\frac{1}{\sqrt{2}}\left(\ket{001} + \ket{110}\right),
\]
which is locally equivalent to the canonical GHZ state under single-qubit basis transformations. Partial quantum state tomography of the dominant density-matrix elements yields a fidelity of $89 \pm 4~\%$ (Fig.~\ref{Figure 3}(c)). The reduced fidelity compared with the two-qubit Bell state reflects the increased sensitivity of three-qubit entanglement to decoherence, as well as the longer pulse sequence required for state preparation.

To quantify the robustness of the generated entangled states, we measure their coherence times by monitoring the decay of the relevant off-diagonal density-matrix elements. The nuclear-spin Bell state exhibits a coherence time of $T_{n_1\text{--}n_2} = 296~ \pm ~21~\upmu\mathrm{s}$ (Fig.~\ref{Figure 3}(b)), which is more than two orders of magnitude longer than the electron-spin dephasing time. In contrast, the coherence time of the three-qubit GHZ-type state is measured to be $T_{\mathrm{GHZ}} = 0.75~ \pm ~0.07~\upmu\mathrm{s}$ (Fig.~\ref{Figure 3}(d)). The shorter lifetime arises from the direct involvement of the electron spin in the entangled state and the cumulative effect of decoherence during the longer preparation sequence, consistent with partial tomography of the off-diagonal coherence terms.

\begin{figure}[htbp]
\centering
\includegraphics[scale = 0.8]{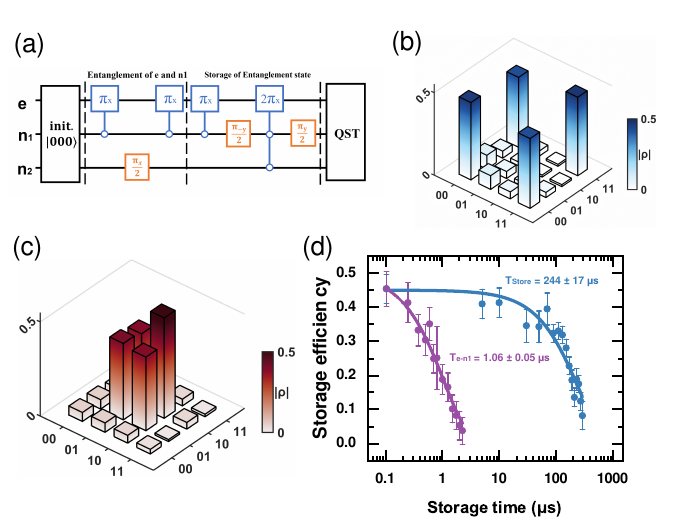}
\caption{Entanglement storage in nuclear-spin memory. 
(a) Schematic of entanglement transfer from the electron–$n_1$ Bell state $(\ket{000}+\ket{110})/\sqrt{2}$ to a nuclear-spin Bell state $(\ket{101}+\ket{110})/\sqrt{2}$ via a sequence of strong-coupling SWAP operations. (b) Reconstructed density matrix of the electron–$n_1$ entangled state. (c) Reconstructed density matrix of the stored nuclear-spin entangled state. (d) Measured coherence decay of the electron–$n_1$ entangled state (purple) and the stored entangled state (blue). The dots represent the raw experimental data, and the solid curves represent the theoretical fits. Error bars represent statistical uncertainties.
}
\label{Figure 4}
\end{figure}

We next demonstrate room-temperature storage of entanglement by coherently transferring an entangled state involving the electron spin onto the nuclear-spin memory qubits. The corresponding pulse sequence is shown in Fig.~\ref{Figure 4}(a). After initialization (init.) of the three-qubit register, we first prepare a Bell state between the electron spin and one nuclear spin while the second nuclear spin remains separable,
\[
\ket{\Psi^+_{e-n_1}}\ket{0}_{n_2}
=
\frac{1}{\sqrt{2}}\left(\ket{000}+\ket{110}\right).
\]
Quantum state tomography (QST) of the electron--nuclear subsystem yields a fidelity of $94.5\pm1.5~\%$ (Fig.~\ref{Figure 4}(b)), confirming high-quality preparation of electron--nuclear entanglement.

To store this entangled state, we apply a sequence of strong-coupling SWAP operations that coherently map the entanglement from the electron--$n_1$ subsystem onto the two nuclear spins $n_1$ and $n_2$, while disentangling the electron spin from the register. The resulting stored state corresponds to the nuclear-spin Bell state (see SM Note 6~\cite{Supplemental} for details)
\[
\ket{\Phi^+_{n_1-n_2}} = \frac{1}{\sqrt{2}}\left(\ket{01} + \ket{10}\right).
\]
Reconstruction of the density matrix after the transfer yields a fidelity of $92.5\pm2.5~\%$ (Fig.~\ref{Figure 4}(c)), demonstrating that the entanglement transfer preserves both population and coherence with minimal loss.

We compare the coherence times of the entangled states before and after storage by monitoring the decay of their off-diagonal density-matrix elements (Fig.~\ref{Figure 4}(d)). The electron-involved entangled state exhibits a coherence time of $1.06 ~\pm ~0.05~\upmu\mathrm{s}$, consistent with the electron-spin coherence measured independently. In contrast, the stored nuclear-spin entangled state displays a dramatically extended lifetime of $244 ~\pm ~17~\upmu\mathrm{s}$, corresponding to an enhancement of more than two orders of magnitude. 

A comparison with the nuclear-spin coherence and electron-spin relaxation times indicates that the storage lifetime is currently influenced by residual electron-mediated decoherence associated with the finite electron-spin $T_1$ relaxation process. Nevertheless, the observed enhancement demonstrates that coherent entanglement transfer enables the preservation of quantum correlations on timescales far exceeding those accessible to the electron spin alone, even under ambient conditions.

To further examine the generality of the entanglement-storage protocol, we applied the same strategy to a distinct heteronuclear register associated with another single shallow PL6 center. This register consists of one strongly coupled $^{29}$Si nuclear spin ($A_{zz}=12.5$~MHz, $\mathrm{Si_{IIa}}$) and one strongly coupled $^{13}$C nuclear spin ($A_{zz}=6.4$~MHz, $\mathrm{C_{II}}$). The heteronuclear register exhibits coherent nuclear-spin control and supports high-fidelity entanglement storage: the initial electron--$^{29}$Si Bell state and the stored $^{29}$Si--$^{13}$C Bell state yield fidelities of $93\pm2~\%$ and $91\pm1.5~\%$, respectively, while the entanglement lifetime is extended from $1.08~\pm~0.09~\upmu$s to $196~\pm~15~\upmu$s. These results, presented in SM Note 5~\cite{Supplemental}, demonstrate that the protocol is not restricted to a specific defect or to a homonuclear $^{29}$Si--$^{29}$Si configuration, but is also applicable to mixed-species nuclear-spin memories with different hyperfine couplings and gyromagnetic ratios.

In addition, a statistical survey of 200 single PL6 centers fabricated using N$_2^+$ and O$^+$ implantation~\cite{Hu2026AdvMater} shows that multi-nuclear-spin registers containing two or more strongly coupled nuclear spins occur with probabilities above 10\%, in good agreement with the theoretical model (see SM Note 4~\cite{Supplemental} for details). This statistical availability indicates that suitable multi-spin resources arise naturally in SiC and are not limited to isolated, exceptionally rare defects.

In conclusion, we demonstrate room-temperature storage of multi-qubit entanglement in silicon carbide by coherently transferring electron--nuclear entanglement to long-lived nuclear-spin qubits. Using a single shallow PL6 center in 4H--SiC, we realize a fully addressable three-qubit register that supports high-fidelity Bell and GHZ-type states and enables a robust SWAP-based entanglement-transfer protocol. The stored entangled state maintains high fidelity and exhibits a lifetime enhanced by more than two orders of magnitude compared with electron-spin-based entanglement.

Beyond the homonuclear $^{29}$Si--$^{29}$Si register, we further validate the protocol in a heteronuclear $^{29}$Si--$^{13}$C register and show that mixed-species nuclear-spin memories can also support high-fidelity entanglement storage. Together with the statistical observation of multi-nuclear-spin registers in PL6 centers created by different implantation methods, these results demonstrate that suitable nuclear-spin memory resources are reproducibly accessible in SiC.

More broadly, our work establishes a practical strategy for shallow defect spin qubits. This electron–nuclear transfer provides a versatile primitive for shallow solid-state spin registers, effectively separating rapid control from long-lived quantum storage and offering a realistic pathway toward entanglement-enhanced quantum sensing under ambient conditions. Leveraging the mature material properties of silicon carbide, our approach also provides a practical route toward scalable quantum devices.

{\bf Acknowledgement:} This work was supported by the National Natural Science Foundation of China (No.\ 92365205, No.\ W2411001, No.\ 62504218, and No.\ 12350006), the Quantum Science and Technology-National Science and Technology Major Project (No.\ 2021ZD0301400 and No.\ 2021ZD0301200), the USTC Major Frontier Research Program (No.\ LS2030000002), USTC (NO.\ YD2030002026) and Shandong Provincial Natural Science Foundation (Grant  No. ZR2024LLZ003). This work was partially performed at the University of Science and Technology of China Center for Micro and Nanoscale Research and Fabrication.

\end{document}